\documentclass[showpacs, pra, aps,  twocolumn,superscriptaddress, 10pt]{revtex4}
\usepackage{mathrsfs}
\usepackage{array}
\usepackage{amsmath}
\usepackage{graphicx}
\usepackage{amstext}
\usepackage{bm}
\usepackage{amsfonts}

\begin{document}
%\title{Robust creation of steady vibrational squeezing and entanglement by optomechanics}
\title{Achieving steady-state entanglement of remote micromechanical oscillators by cascaded cavity coupling}
\author{Huatang Tan}
\email{tanhuatang@phy.ccnu.edu.cn}
\affiliation{Department of physics, Huazhong Normal University, Wuhan 430079, China}
%\affiliation{B2 Institute, Department of Physics and College of Optical Sciences,
%University of Arizona, Tucson, AZ 85721}
\author{L. F. Buchmann}
\affiliation{B2 Institute, Department of Physics and College of Optical Sciences,
University of Arizona, Tucson, AZ 85721}
\author{H. Seok}
\affiliation{B2 Institute, Department of Physics and College of Optical Sciences,
University of Arizona, Tucson, AZ 85721}
\author{Gaoxiang Li}
\affiliation{Department of physics, Huazhong Normal University, Wuhan 430079, China}
%\author{P. Meystre}
%\affiliation{B2 Institute, Department of Physics and College of Optical Sciences,
%University of Arizona, Tucson, AZ 85721}
\begin{abstract}
In this paper, we propose a scheme for generating steady-state entanglement of remote micromechanical oscillators in unidirectionally-coupled cavities. For the system of two mechanical oscillators, we show that when two cavity modes in each cavity are driven at red- and blue-detuned sidebands, respectively, a stationary two-mode squeezed vacuum state of the two mechanical oscillators can be generated with the help of the cavity dissipation. The degree of squeezing is controllable by adjusting the relative strength of the pump lasers. Our calculations also show that the achieved mechanical entanglement is robust against thermal fluctuations of phononic environments.
For the case of multiple mechanical oscillators, we find that the steady-state genuine multipartite entanglement can also be built up among the remote mechanical oscillators by the cavity dissipation. The present scheme does not require nonclassical light input or conditional quantum measurements, and it can be realized with current experimental technology.
\end{abstract}

\maketitle
\section{Introduction}
Besides fundamental research interests in quantum physics \cite{dec}, realizing quantum effects of macroscopic objects is crucial for potential applications in ultrahigh precision measurements and quantum information processing \cite{mech, opm1, opm2}. Thanks to the recent achievements in ground-state cooling of micromechanical oscillators via optomechanical coupling \cite{c2, c3, c4, c5}, the emerging field of cavity optomechanics as an interface between mechanical systems and optical field has become a unique platform to study quantum behavior of macroscopic mechanical systems \cite{app1, app2, app3, app4, app5, app6}. Using well-established quantum optical techniques, optomechanics holds the promise to effectively prepare and manipulate nonclassical mechanical states.

Several schemes have been proposed to establish entanglement between a mechanical element and the driven cavity field \cite{se3} or between vibrating membranes or end mirrors \cite{se1, se2, se4, se5} by  optomechanics. Apart from short-distance mechanical entanglement, remote entanglement between two micromechanical oscillators in separated cavities can also be entangled via injecting squeezed light or conditional quantum measurements \cite{se6, se7, se8}. It was also showed that weak mechanical entanglement between two distant optomechanical oscillators can be possibly achieved merely by optomechanical coupling \cite{se9}. The entanglement of remote mechanical elements is of importance for constructing long-distance quantum communication networks \cite{long}.

On the other hand, generating quantum states by quantum-reservoir engineering has attracted a lot of attention recently. In this approach, the interaction between system and environment is engineered in such a way that the system relaxes into a desired state. The resulting quantum states are steady, independent of initial conditions, and most importantly robust against incoherent noise. To date, several schemes have been proposed to prepare entangled states of atomic systems by quantum dissipation \cite{dis1, re1, re2, re3, re4, re5, re6} and the dissipative creation of steady-state entanglement between two separated cold atomic ensembles has been experimentally realized \cite{dis2}.

In this paper, we consider the generation of steady-state entanglement of remote micromechanical oscillators (membranes) by cavity dissipation. We at first investigate the entanglement between two micromechanical membrane oscillators in a cascaded cavity system. In each cavity, a membrane oscillator is coupled to two nondegenerate cavity modes via parametric and beam-splitter-like interactions by driving the relevant cavity modes on blue- and red-detuned sidebands respectively. For negligible mechanical damping, we find that the cavity dissipation can pull the two distant mechanical oscillators into a stationary two-mode squeezed vacuum. It is also shown that the two-mode entanglement is robust against thermal fluctuations when one takes into account the mechanical damping. We then extend the two-mode mechanical model to the case of multiple mechanical oscillators in an array of cascaded cavities. We show that in this system genuine multipartite steady-state entanglement can be built up among the remote mechanical oscillators with the help of the cavity dissipation.

The reminder of this paper is arranged as follows. In Sec.II, the model of two cascaded optomechanical system is introduced and the steady-state entanglement between the mechanical oscillators is investigated in detail. In Sec.III, we extend the previous model to the case of multiple mechanical oscillators in an array of unidirectionally-coupled cavities and discuss the generation of multipartite entanglement among multiple mechanical oscillators. At last, we give the conclusion in Sec.IV.

\section{Entanglement of two mechanical oscillators}
\subsection{Model and equations}
As schematically shown in Fig.1,  we investigate a system consisting of two identical optical cavities connected by unidirectional coupling \cite{quannoise}. In each cavity, two driven cavity modes are coupled to a vibrating membrane via radiation pressure \cite{mid1, mid2}. The role of the membranes could also be played by other mechanical systems such as trapped clouds of ultracold atoms \cite{cold}. After removing the carrier photons with filters, the output quantum fluctuations from the first cavity are directed to the second cavity to drive the corresponding cavity modes. With the light fields rotating at their driving frequencies, the Hamiltonian of the system is given by
\begin{align}
H/\hbar&=\sum_{j=1,2}\Big[\delta_{a_j}a_j^\dag a_j+\delta_{b_j}b_j^\dag b_j+\omega_{m_j} c_j^\dag c_j\nonumber\\
&~~~~~~~~~+(\tilde{g}_{a_j}a_j^\dag a_j+\tilde{g}_{b_j}b_j^\dag b_j)(c_j+c_j^\dag)\nonumber\\
&~~~~~~~~~+i(\mathcal{E}_{a_j}a_j^\dag-\mathcal{E}_{a_j}^*a_j)+i(\mathcal{E}_{b_j}a_j^\dag-\mathcal{E}_{b_j}^*b_j)\Big]\label{hamil},
\end{align}
where $a_j~(a_j^\dag)$ and $b_j~(b_j^\dag)~(j=1,2)$ are annihilation (creation) operators for the cavity modes and $c_j~(c_j^\dag)$ for the mechanical modes of the vibrating membranes in each cavity.
The cavity-laser detunings $\delta_{z_j}=\omega_{z_j}-\nu_{z_j}~(z=a,b)$, with $\omega_{z_j}$ being the cavity resonant frequencies and $\nu_{z_j}$ the corresponding driving frequencies. The mechanical frequencies of the membranes are denoted by $\omega_{m_j}$ and the optomechanical coupling $\tilde{g}_{z_j}=\sqrt{\hbar/m_j\omega_{m_j}}\omega_{z_j}/L$, with $L$ being the cavity length and $m_j$ the effective mass of the membranes. The amplitudes of the driving lasers $|\mathcal{E}_{z_j}|=\sqrt{2P_{z_j}\tilde{\kappa}_{z_j}/\hbar\nu_{z_j}}$, where $P_{z_j}$ are the powers of the pump lasers and $\tilde{\kappa}_{z_j}$ the cavity loss rates of the left cavity mirrors.
\begin{figure}
\centerline{~~~~~\scalebox{0.45}{\includegraphics{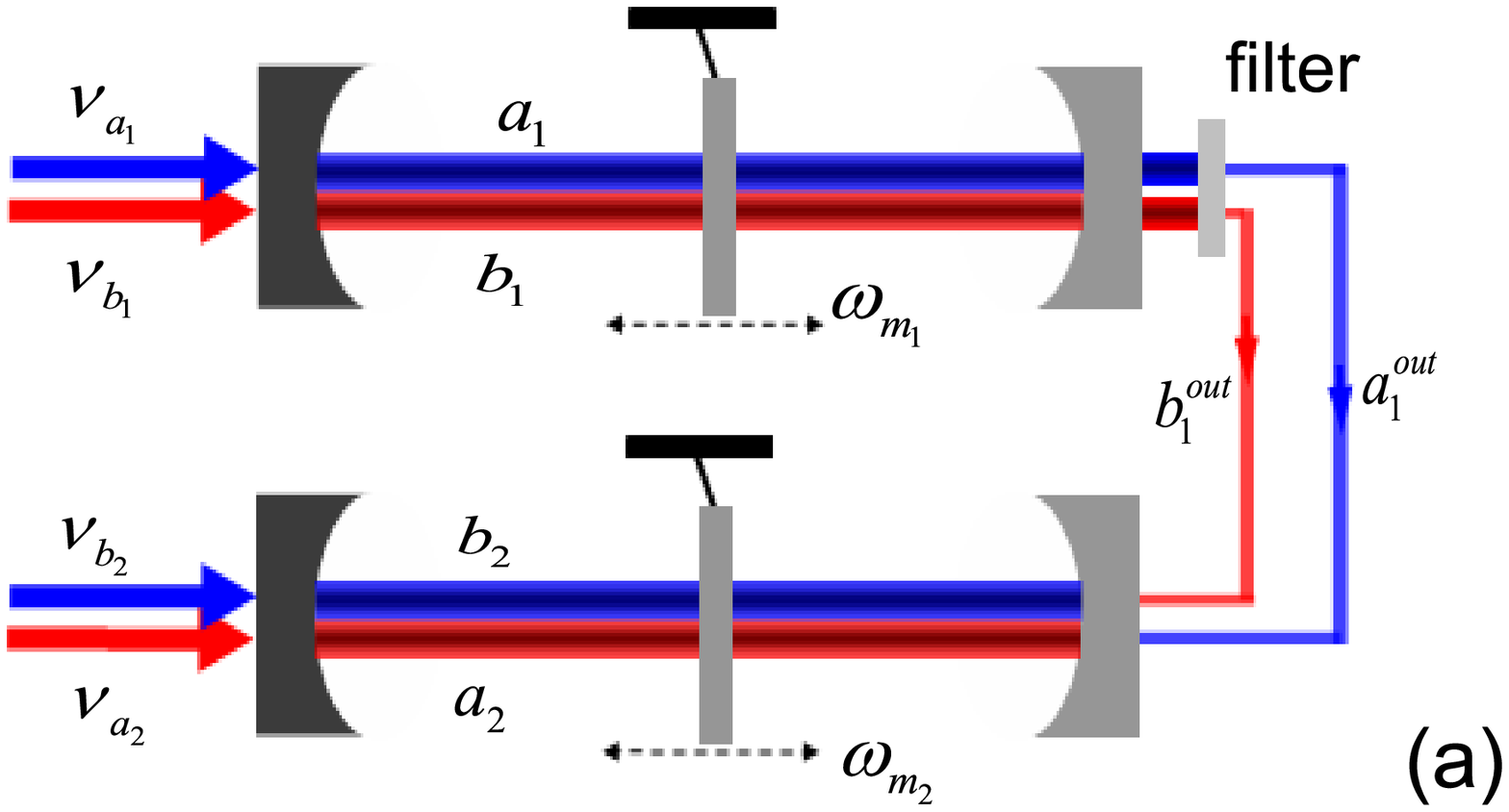}}} \centerline{\scalebox{0.45}{\includegraphics{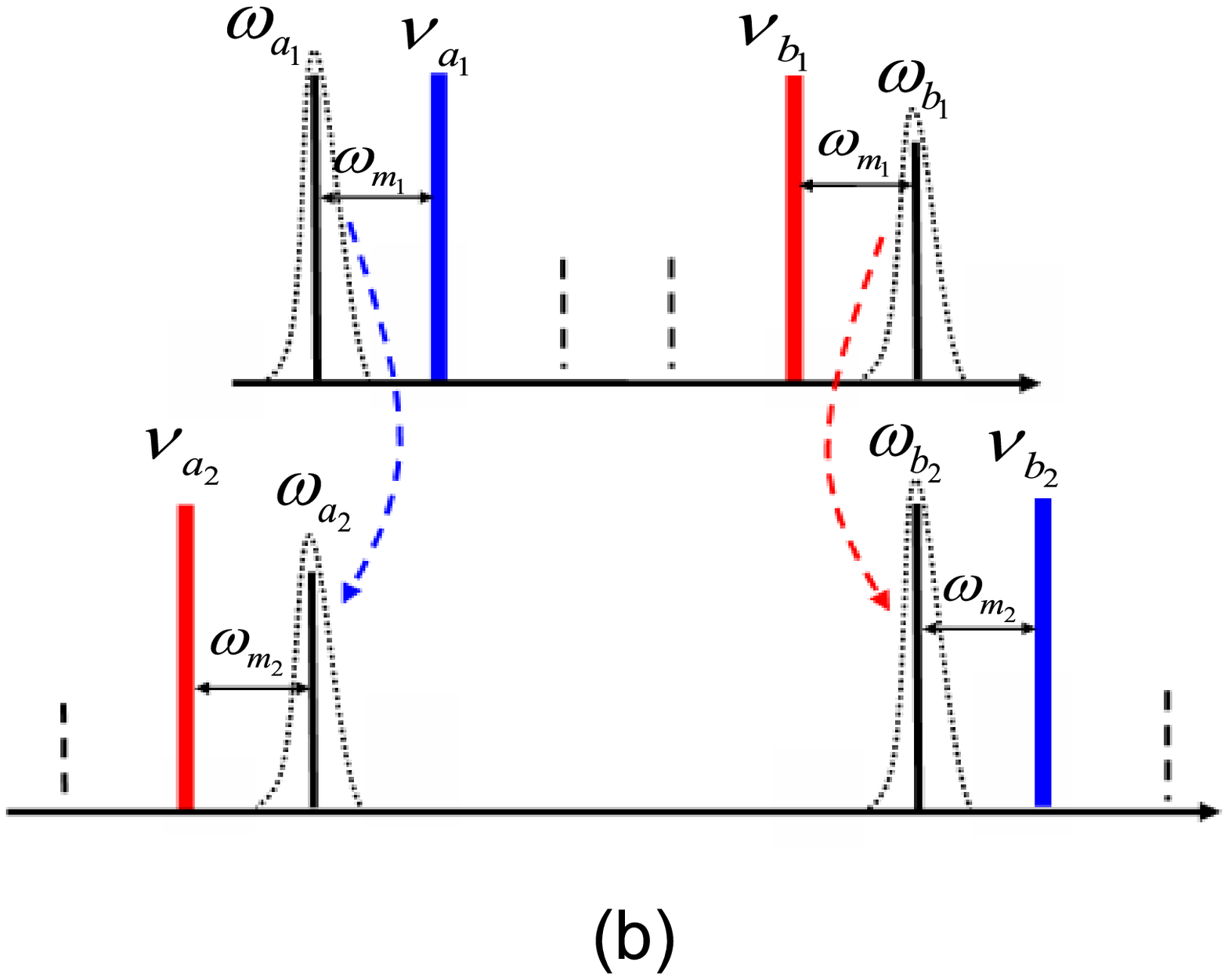}}}
\caption{(a) Schematic setup of two cascaded cavity-optomechanical systems. In each cavity, two cavity modes are driven by red- and blue-detuned lasers, respectively, and the output quantum fluctuations from the first cavity are directed to the second cavity to drive the corresponding cavity modes. (b) Frequencies of the pumps and cavity modes, and the dashed arrows represent the unidirectional coupling between the cavity modes.}
\end{figure}

We expand the quantum operators as $o_j=\bar{o}_j^s+\delta o_j$, where $\bar{o}_j^s$ are the steady-state classical amplitudes and $\delta o_j$ the corresponding quantum fluctuation operators. By taking into account cavity losses and mechanical damping, the classical amplitudes are obtained as $\bar{z}_j^s=\mathcal{E}_{z_j}/(\kappa_{z_j}+i\Delta_{z_j})$ and $\bar{c}_j^s=\sum_z \tilde{g}_{z_j}|\bar{z}_j^s|^2/(\omega_{m_j}+i\gamma_{m_j})$, where $\Delta_{z_j}=\delta_{z_j}+2\tilde{g}_{z_j}\text{Re}(\bar{c}_j^s)$, $\kappa_{z_j}$ are the cavity loss rates from the output mirrors on the right of the cavities, and $\gamma_{m_j}$ are the mechanical damping rates. Note that here we have assumed the cavity loss rates $\kappa_{z_j}\gg \tilde{\kappa}_{z_j}$ such that losses from the left cavity mirrors can be neglected. For intense driving fields we have $|\bar{o}_j^s|^2\gg \langle \delta o_j^\dag \delta o_j\rangle$ and the Hamiltonian (\ref{hamil}) can be linearized. Then, by dropping the symbol $``\delta"$ in the fluctuation operators for simplicity of notation, the resulting Langevin equations of motion for the quantum fluctuations of the cavity and mechanical modes are obtained as
\begin{align}
\dot{a}_j=&-(\kappa_{a_j}+i\Delta_{a_j})a_j-ig_{a_j}(c_j+c_j^\dag)+\sqrt{2\kappa_{a_j}}a_{j}^{\text{in}}(t),\nonumber\\
\dot{b}_j=&-(\kappa_{b_j}+i\Delta_{b_j})b_j-ig_{b_j}(c_j+c_j^\dag)+\sqrt{2\kappa_{b_j}}b_{j}^{\text{in}}(t),\nonumber\\
\dot{c}_j=&-(\gamma_{m_j}+i\omega_{m_j})c_j-ig_{a_j}(a_j+a_j^\dag)-ig_{b_j}(b_j+b_j^\dag)\nonumber\\
&+\sqrt{2\gamma_{m_j}}c_{j}^{\text{in}}(t),\label{lan1}
\end{align}
where the effective optomechanical coupling $g_{z_j}=|\bar{z}_j^s| \tilde{g}_{a_j}~(z=a,b)$. The noise operators $a_1^{\text{in}}(t)$ and $b_1^{\text{in}}(t)$ describe vacuum inputs to the first cavity and satisfy nonzero correlations $\langle a_1^{\text{in}}(t)a_1^{\text{in}\dag}(t')\rangle=\delta(t-t')$ and $\langle b_1^{\text{in}}(t)b_1^{\text{in}\dag}(t')\rangle=\delta(t-t')$. The input noise of the second cavity, characterized by the operators $a_2^{\text{in}}(t)$ and $b_2^{\text{in}}(t)$, are from the output fluctuations of the first cavity and transmission losses in the coupling. When the output quantum field of the cavity mode $a_1~(b_1)$ is used to drive the cavity mode $a_2~(b_2)$, then one has
\begin{subequations}
\begin{align}
a_2^{\text{in}}(t)&=\sqrt{\eta_a}[a_1^{\text{in}}(t)-\sqrt{2\kappa_{a_1}}a_1(t)]e^{-i(\nu_{a_1}-\nu_{a_2})t}\nonumber\\
&~~+\sqrt{(1-\eta_a)}\tilde{a}_2^{\text{in}}(t),\\
b_2^{\text{in}}(t)&=\sqrt{\eta_b}[b_1^{\text{in}}(t)-\sqrt{2\kappa_{b_1}}b_1(t)]e^{-i(\nu_{b_1}-\nu_{b_2})t}\nonumber\\
&~~+\sqrt{(1-\eta_b)}\tilde{b}_2^{\text{in}}(t),
\end{align}
\end{subequations}
where $\eta_z\in[0,1]~(z=a,b)$ accounts for the imperfect couplings between the two cavities. The operators $\tilde{a}_2^{\text{in}}(t)$ and $\tilde{b}_2^{\text{in}}(t)$ denote the local vacuum noise input to the second cavity. The parameter $\eta_z=1$ corresponds to a lossless unidirectional coupling between the two cavities, whereas $\eta_z=0$ describes two independent cavities.  Note here that the exponential factors in the above equations result from the differences between the frequencies of the relevant pump lasers. In addition, $c_j^{\text{in}}(t)$ are noise operators of the mechanical oscillators which have nonzero correlations  $\langle c_j^{\text{in}\dag}(t)c_j^{\text{in}}(t')\rangle=\bar{n}_{\text{th}}^j\delta(t-t')$ and $\langle c_j^{\text{in}}(t)c_j^{\text{in}\dag}(t')\rangle=(\bar{n}_{\text{th}}^j+1)\delta(t-t')$, where the mean thermal phonon numbers at temperature $T$ is given by $\bar{n}_{\text{th}}^j=(e^{\hbar\omega_{m_j}/k_BT}-1)^{-1}$, with $k_B$ the Boltzmann constant.

Now we choose the detunings
\begin{align}
\Delta_{a_1}=-\Delta_{b_1}=-\omega_{m_1},~\Delta_{a_2}=-\Delta_{b_2}=\omega_{m_2},
\end{align}
i.e., the cavity modes $a_1$ and $b_2$ are pumped by lasers which are blue-detuned from their resonance frequencies by the mechanical frequencies, while the modes $a_2$ and $b_1$ are driven by pump lasers which are red detuned by the same amount, as illustrated in Fig. 1(b). Therefore, the pump frequencies $\nu_{x_j}$ should satisfy
\begin{subequations}
\begin{align}
\nu_{a_1}-\nu_{a_2}&=(\omega_{m_1}+\omega_{m_2}), \\
\nu_{b_1}-\nu_{b_2}&=-(\omega_{m_1}+\omega_{m_2}).
\end{align}
\end{subequations}
With the above choices of detunings, by
performing the transformations $z_j\rightarrow z_je^{-i\Delta_{z_j}t}$, $z_j^{\text{in}}(t)\rightarrow z_j^{\text{in}}(t)e^{-i\Delta_{z_j}t}~(z=a,b)$, $c_j\rightarrow c_je^{-i\omega_{m_j}t}$, and $c_j^{\text{in}}(t)\rightarrow c_j^{\text{in}}(t)e^{-i\omega_{m_j}t}$, and neglecting fast oscillating terms proportional to $e^{\pm i(\omega_{m_1}+\omega_{m_2})t}$, the Langevin equations (\ref{lan1}) reduce to
\begin{subequations}
\label{lan2}
\begin{align}
&\dot{a}_1=-\kappa_{a_1}a_1-ig_{a_1}c_1^\dag+\sqrt{2\kappa_{a_1}}a_{1}^{\text{in}}(t),\\
&\dot{b}_1=-\kappa_{b_1}b_1-ig_{b_1}c_1+\sqrt{2\kappa_{b_1}}b_{1}^{\text{in}}(t),\\
&\dot{a}_2=-\kappa_{a_2}a_2-ig_{a_2}c_2-2\sqrt{\eta_a\kappa_{a_1}\kappa_{a_2}}a_1
+\sqrt{2\eta_a\kappa_{a_2}}a_{1}^{\text{in}}(t)\nonumber\\
&~~~~~~+\sqrt{2(1-\eta_a)\kappa_{a_2}}\tilde{a}_2^{\text{in}}(t),\\
&\dot{b}_2=-\kappa_{b_2}b_2-ig_{b_2}c_2^\dag-2\sqrt{\eta_b\kappa_{b_1}\kappa_{b_2}}b_1
+\sqrt{2\eta_b\kappa_{b_2}}b_{1}^{\text{in}}(t)\nonumber\\
&~~~~~~+\sqrt{2(1-\eta_b)\kappa_{b_2}}\tilde{b}_2^{\text{in}}(t),\\
&\dot{c}_1=-\gamma_{m_1}c_1-ig_{a_1}a_1^\dag-ig_{b_1}b_1
+\sqrt{2\gamma_{m_1}}c_{1}^{\text{in}}(t),\\
&\dot{c}_2=-\gamma_{m_2}c_2-ig_{a_2}a_2-ig_{b_2}b_2^\dag
+\sqrt{2\gamma_{m_2}}c_{2}^{\text{in}}(t).
\end{align}
\end{subequations}
It should be noted that for our approximations to be valid, we require our system to be in the resolved sideband regime, $\omega_{m_j}\gg \kappa_{z_j}$, as well as to satisfy $\omega_{m_j}\gg g_{z_j}$.
The above equations show that in each cavity, the mechanical mode is coupled to the cavity modes via effective parametric amplification as well as beam-splitter-like mixing. While the former interaction leads to photon-phonon entanglement and optical amplification, the latter is damping the mechanical modes. If the coupling strengths satisfy $g_{b_1}>g_{a_1}$ and $g_{a_2}>g_{b_2}$, optical damping is dominant over amplification and both mechanical oscillators are cooled.

\subsection{Two-mode mechanical entanglement}
We can equivalently reexpress Eqs. (\ref{lan2}) as $\dot{\chi}=\mathcal{Z}\chi+f^{\text{in}}(t)$, with the vector $\chi=(x_{a_1}, y_{a_1},x_{b_1}, y_{b_1}, x_{a_2}, y_{a_2}, x_{b_2}, y_{b_2}, x_{c_1}, y_{c_1}, x_{c_2}, y_{c_2})^T$, in terms of the quadrature operators defined as $x=(o+o^\dag)/\sqrt{2}$ and $y=-i(o-o^\dag)/\sqrt{2}$, while $f^{\text{in}}(t)$ contains the corresponding noise operator contributions. The entanglement between the mechanical systems is contained in the $12\times12$ correlation matrix $\tilde{\sigma}$ given by $\tilde{\sigma}_{ij}=\langle \chi_i\chi_j+\chi_j\chi_i\rangle/2$. In steady-steady state, it satisfies $\mathcal{Z}\tilde{\sigma}_s+\tilde{\sigma}_s\mathcal{Z}^T=-\mathcal{D}$, where $D$ is the noise matrix $\mathcal{D}_{ij}\delta(t-t')=\langle f_i^{\text{in}}(t)f_j^{\text{in}}(t')+f_j^{\text{in}}(t')f_i^{\text{in}}(t)\rangle/2$. Since we are only interested in the entanglement between the two mechanical modes, it is enough to consider the reduced correlation matrix $\sigma_{12}$ related to the two-mode mechanical states. It has the simple structure
$\sigma_{12}=\begin{pmatrix} \sigma_{12}^1 & \sigma_{12}^3 \\
(\sigma_{12}^3)^T & \sigma_{12}^2\end{pmatrix}$, where $\sigma_{12}^1$, $\sigma_{12}^2$, and $\sigma_{12}^3$ are $2\times2$ matrices containing the autocorrelations of the two systems and their cross-correlations respectively.
The entanglement between the two mechanical modes can be quantified with the logarithmic negativity $E_{12}$ \cite{En}, which is defined as
\begin{align}
E_{12}=\text {max}[0, -\ln(2\zeta_{12})],\label{log}
\end{align}
where $\zeta_{12}$ is given in terms or the reduced correlation matrix
\begin{equation}
\zeta_{12}=2^{-1/2}\sqrt{\Sigma(\sigma_{12})-\sqrt{\Sigma(\sigma_{12})-4\text{det}\sigma_{12}}},
\end{equation}
 with $\Sigma(\sigma_{12})=\text {det}\sigma_{12}^1+\text {det} \sigma_{12}^2-2\text{det} \sigma_{12}^3$.

Solving Eqs.(\ref{lan2}) numerically and using Eq. (\ref{log}) we can investigate the mechanical entanglement in the system. Let us first, however, turn to a regime where we can obtain analytical results. To this end, we consider the cavity dissipation rates $\kappa_{z_j}=\kappa$, the perfect cavity couplings $\eta_z=1$, and the effective optomechanical couplings
\begin{align}
g_{a_1}=g_{b_2}=g_1,~~g_{a_2}=g_{b_1}=g_2.
\end{align}
If the cavity dissipation rate is dominating the dynamics of the system, i.e. $\kappa\gg\{g_j,\gamma_{m_j}\bar{n}_{\text{th}}^j\}$, the cavity modes follow changes of the mechanical oscillators adiabatically for times $t>1/\kappa$.
%\begin{align}
%\frac{d}{dt}c_1(t)&=-[\gamma_1
%+(\frac{g_1^2}{\kappa}-\frac{g_2^2}{\kappa})]c_1+\sqrt{2\gamma_1}c_1^{\text{in}}(t)
%\end{align}
In this case we can eliminate the cavity modes and find the simple equations of motion for the mechanical modes $c_j$
\begin{subequations}
\label{lan3}
\begin{align}
\dot{c}_1(t)=-(\gamma_{m_1}+\tilde{\gamma}_m)c_1(t)+\sqrt{2\gamma_{m_1}}c_1^{\text{in}}(t)+\tilde{c}_1^{\text{in}}(t),\\
\dot{c}_2(t)=-(\gamma_{m_2}+\tilde{\gamma}_m)c_2(t)+\sqrt{2\gamma_{m_2}}c_2^{\text{in}}(t)+\tilde{c}_2^{\text{in}}(t),
\end{align}
\end{subequations}
where $\tilde{\gamma}_m=(g_2^2-g_1^2)/\kappa$ is the net optomechanical damping rate. The noise operators $\tilde{c}_j^{\text{in}}(t)$ are given by
\begin{subequations}
\begin{align}
\label{no}
\tilde{c}_1^{\text{in}}(t)=-\frac{\sqrt{2}ig_1}{\sqrt{\kappa}}a_1^{\text{in}\dag}(t)
-\frac{\sqrt{2}ig_2}{\sqrt{\kappa}}b_1^{\text{in}}(t),\\
\tilde{c}_2^{\text{in}}(t)=\frac{\sqrt{2}ig_2}{\sqrt{\kappa}}a_1^{\text{in}}(t)
+\frac{\sqrt{2}ig_1}{\sqrt{\kappa}}b_1^{\text{in}\dag}(t),
\end{align}
\end{subequations}
and have non-vanishing correlations
\begin{subequations}
\label{cor}
\begin{gather}
\langle\tilde{c}_j^{\text{in}\dag}(t)\tilde{c}_j^{\text{in}}(t')\rangle=2\tilde{\gamma}_mN_m\delta(t-t'),\\
\langle\tilde{c}_j^{\text{in}}(t)\tilde{c}_j^{\text{in}\dag}(t')\rangle=2\tilde{\gamma}_m(N_m+1)\delta(t-t'),\\
\langle\tilde{c}_1^{\text{in}}(t)\tilde{c}_2^{\text{in}}(t')\rangle=2\tilde{\gamma}_m\sqrt{N_m(N_m+1)}\delta(t-t'),
\end{gather}
\end{subequations}
with $N_m=g_1^2/(g_2^2-g_1^2)$. The above correlations indicate that the two mechanical oscillators are effectively coupled to a broadband quantum reservoir in a two-mode squeezed vacuum state \cite{qo}.
In the absence of the mechanical damping ($\gamma_{m_j}=0$), the mechanical oscillators will reduce to the state of the reservoir in the long-time limit, i.e., the two-mode squeezed vacuum
\begin{align}
|\psi\rangle_{12}^{ss}=\exp (-rc_1^\dag c_2^\dag+r c_1c_2)|0_{c_1},0_{c_2}\rangle,
\end{align}
with the squeezing parameter  $r=\tanh^{-1}(g_1/g_2)$ only dependent on the relative strengths of the two pump lasers. Therefore, the strong mechanical entanglement can be built up in principle just by controlling the ratio of the strengths of the pump lasers.
It should be pointed that our scheme is quite different from that in Ref.\cite{se6} which discussed the establishment of the stationary entanglement between two mechanical oscillators by injecting externally squeezed light into the cavities. Here, instead of creating entanglement in an external source, the entanglement between the mechanical oscillator and blue-detuned cavity mode is created via the parametric interaction in each cavity, and the photon-phonon entanglement is then transferred to the mechanical oscillators with the help of the beam-splitter interaction.
\begin{figure}
\centerline{\scalebox{0.3}{\includegraphics{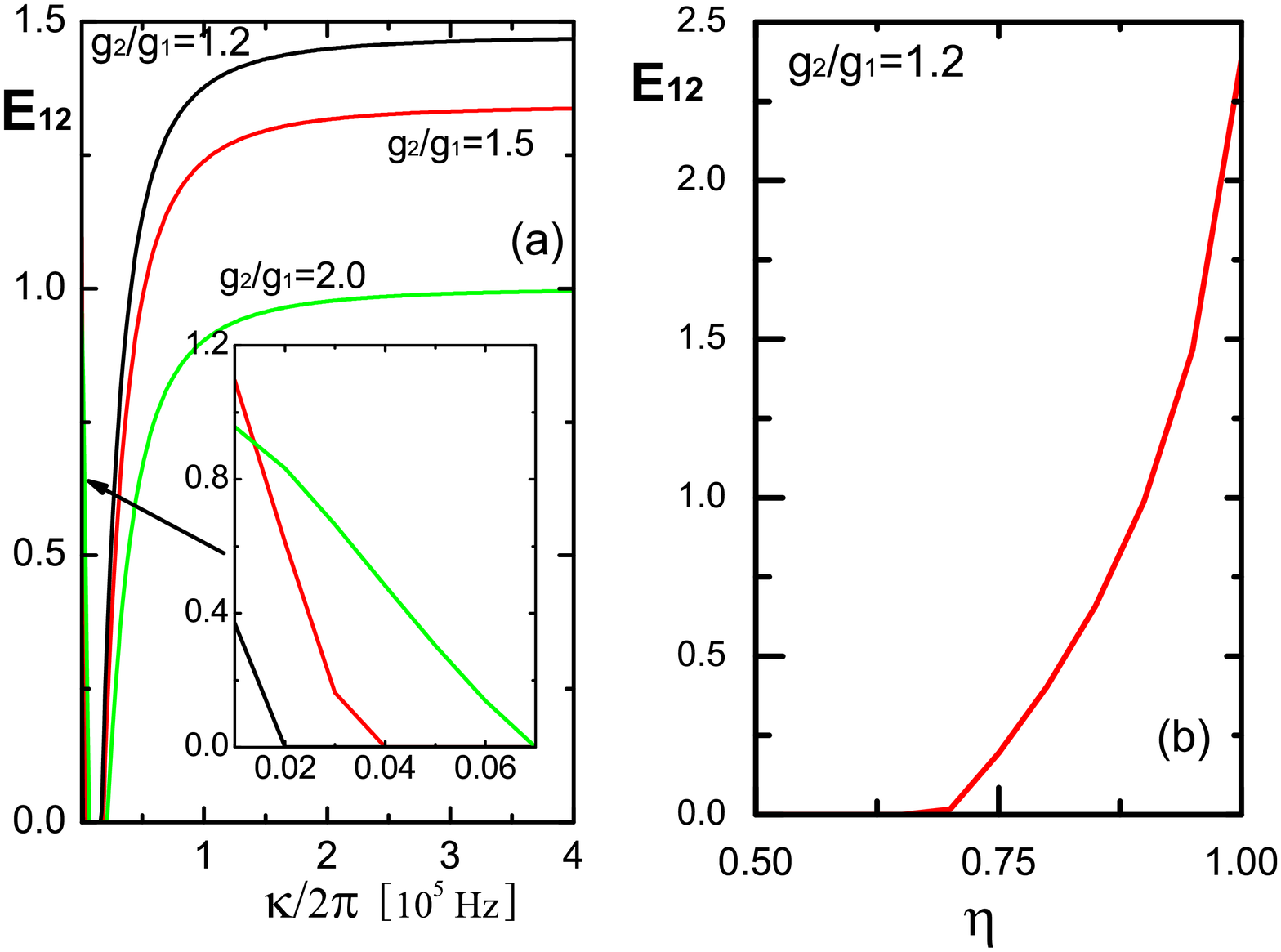}}} \caption{(a) Dependence of the steady-state mechanical entanglement $E_{12}$ on the cavity dissipation rate $\kappa$ for different coupling ratios $g_2/g_1$. Other parameters are the mechanical decay rate $\gamma_m=0$, the coupling strength $g_1/2\pi=0.1\times 10^5$ Hz, and the unidirectional intercavity coupling efficiency $\eta=0.95$. (b) The mechanical entanglement as a function of coupling efficiency $\eta$ for the relative strength $g_2/g_1=1.2$ and the other parameters are the same as in (a).}
\end{figure}
\begin{figure}
\centerline{\scalebox{0.3}{\includegraphics{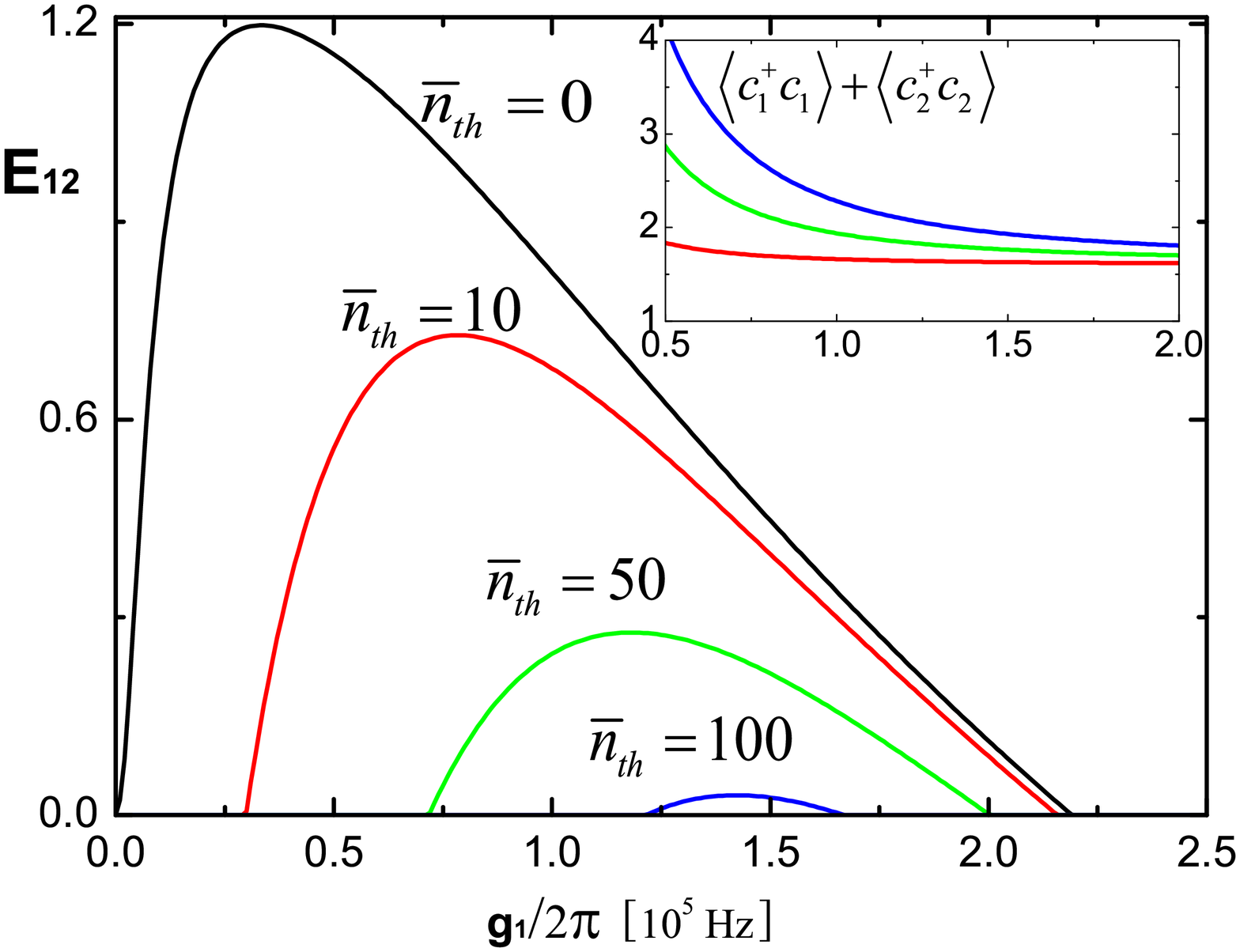}}} \caption{Dependence of the mechanical entanglement on the coupling $g_1$ for different values of mean thermal phonon number $\bar{n}_{\text{th}}$, with the cavity decay rate $\kappa/2\pi=4\times 10^5$ Hz, the mechanical damping rate $\gamma_m/2\pi=100$ Hz, the
coupling $g_2=1.5g_1$, and unidirectional coupling efficiency $\eta=0.95$.
%The inset gives the corresponding mean phonon numbers of the two mechanical modes.
%Parameter values are $\kappa/2\pi\simeq 4\times 10^5$Hz, $\omega_m/2\pi\simeq 2\times10^6$Hz,
%$\gamma_m/2\pi\simeq 100$Hz, and $g_1/2\pi\simeq(0.2-1.5)\times 10^5$Hz.
}
\end{figure}

By taking into account mechanical damping, from Eq.(\ref{lan3}) we have the steady-state values $
\langle c_j^\dag c_j\rangle=(\gamma_m\bar{n}_{\text{th}}+\tilde{\gamma}_mN_m)/(\gamma_m+\tilde{\gamma}_m)$ and
$\langle c_1 c_2\rangle=\tilde{\gamma}_mM_m/(\gamma_m+\tilde{\gamma}_m)$, where we have assumed $\gamma_{m_1}=\gamma_{m_2}\equiv\gamma_m$ and $\bar{n}_{\text{th}}^1=\bar{n}_{\text{th}}^2\equiv\bar{n}_{\text{th}}$ for simplicity. It is easy to the entanglement parameter
\begin{align}
\zeta_{12}=\frac{1}{2}-\frac{g_1(g_2-g_1)-\kappa\gamma_m\bar{n}_{\text{th}}}{\kappa\gamma_m+g_2^2-g_1^2}.
\end{align}
Clearly, steady-state mechanical entanglement can be achieved at non-zero temperature, provided that the mean number of thermal phonons satisfy
\begin{equation}
\bar{n}_{\text{th}}<\frac{g_1g_2}{\kappa\gamma_m}(1-\frac{g_1}{g_2}).
\end{equation}
Given that the couplings $g_j$ are tunable through the pump lasers, this condition demonstrates the robustness of steady-state entanglement against thermal noise in the mechanical systems.

We next turn to the numerical results from solving Eqs.(\ref{lan2}), which allows us to investigate the entanglement property in the regime where the adiabatical elimination of the cavity modes is invalid. In Fig.2 the dependence of steady-state mechanical entanglement on the cavity decay rate $\kappa$ is plotted for different values of $g_2/g_1$ and the mechanical damping $\gamma_m=0$. Consistently with our analytic results we observe for large cavity decay $\kappa\gg g_j$, the entanglement becomes saturated and independent of $\kappa$. The increase of the entanglement with decreasing coupling ratios $g_2/g_1$ is also evident in this regime. Furthermore, for the converse situation $\kappa\ll g_j$, we also observe the steady-state entanglement, although to a smaller degree than in the adiabatic regime. The behavior of the steady-state entanglement in the presence of mechanical damping is demonstrated in Fig.3.
We see that in this case the optimal entanglement does not occur in the adiabatical regime. With increasing thermal phonon number $\bar{n}_{\text{th}}$ stronger coupling strengths $g_j$ are needed to achieve the maximum entanglement. However, the robustness of the generated entanglement is obvious,  as it can still be maintained for a relatively high mean thermal phonon number $\bar{n}_{\text{th}}=100$. Reaching the quantum ground-state of the vibrational modes is therefore not a prerequisite of the present scheme, which reduces experimental difficulties considerably.

\section{Multipartite mechanical entanglement}
\begin{figure}
\centerline{\scalebox{0.28}{\includegraphics{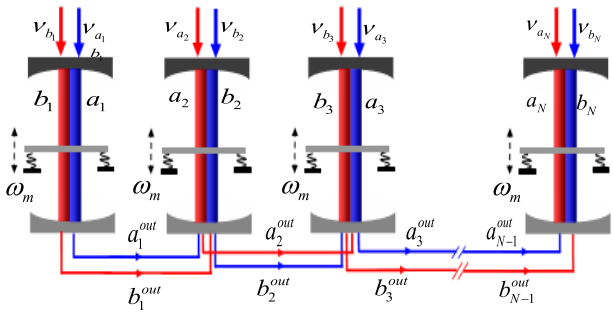}}} \caption{Schematic plot of $N$ vibrating membranes trapped in cascaded cavities.}
\end{figure}

In this section, we generalize the previous two-cavity model to a system of $N$ mechanical oscillators in coupled cavities, as illustrated in Fig. 4 and proceed to discuss the generation of multipartite mechanical entanglement.
Staying consistent with the previous model, we introduce the convention that the cavity modes $a_{2n-1}$ and $b_{2n}$ are pumped by lasers which are blue-detuned from their resonances, while $a_{2n}$ and $b_{2n-1}$ are driven by red detuned pumps. Therefore, the red and blue detuned modes are $a_{2n-1}$ and $b_{2n-1}$ respectively for odd cavities, while they are $b_{2n}$ and $a_{2n}$ for even cavities, see Fig. 1.(b). Assuming identical mechanical frequencies for all oscillators ($\omega_m$) and identical cavities, we thus find the effective detunings
\begin{align}
\Delta_{a_{2n-1}}=-\Delta_{b_{2n-1}}=-\omega_m,~~\Delta_{a_{2n}}=-\Delta_{b_{2n}}=\omega_m,
\end{align}
and for the driving frequencies
\begin{align}
\nu_{a_{2n}}-\nu_{a_{2n-1}}=-2\omega_m,~~\nu_{b_{2n}}-\nu_{b_{2n-1}}=2\omega_m.
\end{align}
With the same procedures and approximations as before, the Langevin equations of motion for the cavity modes $z_j$ and the mechanical modes $c_j$ can be obtained and read
\begin{subequations}
\label{nonadiab:lan4}
\begin{align}
\dot{z}_{j}=&-\kappa_z z_{j}-ig_z c_j^z-2\kappa_z\sum_{s=1}^{j-1}(\sqrt{\eta_z})^{j-s}z_s\nonumber\\
&+\sqrt{2\kappa_z}\sum_{s=2}^j\sqrt{\eta_z^{j-s}(1-\eta_z)}\tilde{z}_s^{\text{in}}(t)\nonumber\\
&+\sqrt{2\kappa_z}(\sqrt{\eta_z})^{j-1}z_{1}^{\text{in}}(t),\\
%\dot{a}_{2j-1}=&-\kappa_a a_{2j-1}-ig_1 c_{2j-1}^\dag-2\kappa_a\sum_{s=1}^{2j-2}(\sqrt{\eta_a})^{2j-1-s}a_s\nonumber\\
%&+\sqrt{2\kappa_a}\sum_{s=2}^{2j-1}\sqrt{\eta_a^{2j-1-s}(1-\eta_a)}\tilde{a}_s^{\text{in}}(t)\nonumber\\
%&+\sqrt{2\kappa_a}(\sqrt{\eta_a})^{2j-2}a_{1}^{\text{in}}(t)\\
%\dot{b}_{2j-1}=&-\kappa_bb_{2j-1}-ig_2c_{2j-1}-2\kappa_b\sum_{s=1}^{2j-2}(\sqrt{\eta_b})^{2j-1-s}b_s\nonumber\\
%&+\sqrt{2\kappa_b}(\sqrt{\eta_b})^{2j-2}b_{1}^{\text{in}}(t),\\
%\dot{a}_{2j}=&-\kappa_aa_{2j}-ig_2c_{2j}-2\kappa_a\sum_{s=1}^{2j-1}(\sqrt{\eta_a})^{2j-s}a_s\nonumber\\
%&+s\sqrt{2\kappa_a}\sum_{s=1}^{2j-1}(\sqrt{\eta_a})^{2j-s}a_s\nonumber\\
%&+\sqrt{2\kappa_a}(\sqrt{\eta_a})^{2j-1}a_{1}^{\text{in}}(t),\\
%\dot{b}_{2j}=&-\kappa_{b_2}b_2-ig_1c_{2j}^
%\dag-2\kappa_b\sum_{s=1}^{2j-1}(\sqrt{\eta_b})^{2j-s}b_s\nonumber\\
%&+\sqrt{2\kappa_b}(\sqrt{\eta_b})^{2j-1}b_{1}^{\text{in}}(t),\\
%\dot{c}_{2j-1}=&-\gamma_mc_{2j-1}-ig_1 a_{2j-1}^\dag-ig_2 b_{2j-1}
%+\sqrt{2\gamma_m}c_{1}^{\text{in}}(t),\nonumber\\
%\dot{c}_{2j}=&-\gamma_mc_{2j}-ig_2a_{2j}-ig_1b_{2j}^\dag
%+\sqrt{2\gamma_m}c_{2}^{\text{in}}(t).\\
\dot{c}_{j}=&-\gamma_mc_j-ig_a a_j^x-ig_b b_j^x
+\sqrt{2\gamma_m}c_j^{\text{in}}(t),
\end{align}
\end{subequations}
where the symbols are
\begin{equation*}
(g_a, g_b, c_j^a, c_j^b, a_j^x, b_j^x)=\left\{
\begin{array}{lr}
(g_1, g_2, c_j^\dag, c_j, a_j^\dag, b_j) & \textrm{for } j \textrm{ odd}\\
(g_2, g_1, c_j, c_j^\dag, a_j, b_j^\dag) & \textrm{for } j \textrm{ even}
\end{array}
\right.
\end{equation*}
with matched optomechanical couplings $g_{a_{2n-1}}=g_{b_{2n}}=g_1$ and $g_{b_{2n-1}}=g_{a_{2n}}=g_2$, local vacuum noise operators $\tilde{z}_{s}(t)$, optical and mechanical loss-rates designated by $\kappa_z$ and $\gamma_m$ respectively and finally the coupling efficiencies $\eta_z$.
Before we turn to the numerical solutions, let us first consider the situation that $\kappa_z\gg\{g_j,\gamma_m\bar{n}_{\text{th}}\}$, which allows us to adiabatically eliminate the cavity modes. For the perfect intercavity couplings $\eta_z=1$ and identical cavity loss rates $\kappa_z=\kappa$, the equations of motion for the odd and even mechanical oscillators are
\begin{subequations}
\label{adiab:lan4}
\begin{align}
\dot{c}_{2n-1}(t)&=-(\gamma_m+\tilde{\gamma}_m)c_{2n-1}(t)-2\tilde{\gamma}_m\sum_{s=1}^{n-1}c_{2s-1}(t)\nonumber\\
&~~~~+\sqrt{2\gamma_m}c_{2n-1}^{\text{in}}(t)+\tilde{c}_1^{\text{in}}(t),\\
\dot{c}_{2n}(t)&=-(\gamma_m+\tilde{\gamma}_m)c_{2n}(t)-2\tilde{\gamma}_m\sum_{s=1}^{n-1}c_{2s}(t)\nonumber\\
&~~~~+\sqrt{2\gamma_m}c_{2n}^{\text{in}}(t)+\tilde{c}_2^{\text{in}}(t).
\end{align}
\end{subequations}
From the above equations we see that the odd and even mechanical oscillators are coupled to the noise operators $\tilde{c}_1^{\text{in}}(t)$ and $\tilde{c}_2^{\text{in}}(t)$, respectively. Therefore, the entanglement may be established between any odd and even mechanical oscillators with the nonclassical correlations between the noises $\tilde{c}_1^{\text{in}}(t)$ and $\tilde{c}_2^{\text{in}}(t)$ given in Eq.(\ref{cor}). However, between oscillators with same parity, quantum entanglement cannot be established. This because that the source of entanglement in this scheme results from the coupling of the red sideband output into the blue sideband input and vice versa. For two even or odd oscillators, the cavity modes coupled to these two oscillators have the same detunings from the pump lasers, which leads them not to being entangled but to the mode coupling through an incoherent exchange interaction with rate $-2\tilde{\gamma}_m$. These results are verified in the following via numerical solution of Eqs.(\ref{nonadiab:lan4}).

\begin{figure}
\centerline{\scalebox{0.3}{\includegraphics{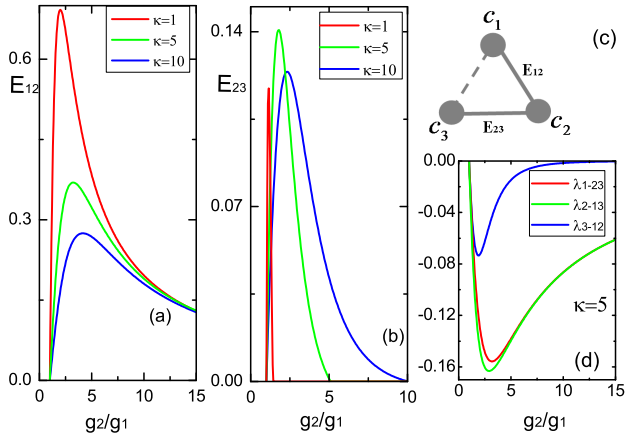}}} \caption{The steady-state reduced bipartite entanglement $E_{12}$ between the mechanical modes $c_1$ and $c_2$ in (a) and the reduced bipartite entanglement $E_{23}$ between the modes $c_2$ and $c_3$ in (b) for a three-mode mechanical system. The tripartite entanglement structure of the  mechanical oscillators is shown in (c) where the solid lines represent the reduced bipartite entanglement. The plot in (d) depicts the negative eigenvalues of the partially transposed correlation matrix of the three-mode mechanical system with respect to any one mechanical mode, which demonstrates that the genuine tripartite mechanical entanglement can be achieved. The unit of $\kappa$ is $10^5$ Hz, the coupling strength $g_1/2\pi=0.01\times 10^5$ Hz, the mechanical rate $\gamma_m/2\pi=10$ Hz, the mean thermal phonon number $\bar{n}_{\text{th}}=0$, and the intercavity coupling efficiency $\eta=1.0$.}
\end{figure}

For $N=3$, we plot in Fig.5 (a) and (b) the bipartite entanglement $E_{12}$ between the mechanical modes $c_1$ and $c_2$ and the bipartite entanglement $E_{23}$ of the modes $c_2$ and $c_3$, respectively. We see that the entanglement $E_{12}>E_{23}$ for the same parameters. As predicted above, bipartite entanglement between the mechanical modes $c_1$ and $c_3$ is absent. Nevertheless, as demonstrated in Fig.5 (d), full inseparable (genuine) tripartite entanglement can be established among the three remote mechanical oscillators. Fig.5~(d) depicts the negative eigenvalues $\lambda_{l-mn}$ of the partially transposed three-mode correlation matrix with respect to the $l$-th mode. The appearance of negative eigenvalue confirms bipartite entanglement between the transposed mode $l$ and the subsystem of the remaining modes $m$ and $n$, and fully inseparable (genuine) multipartite entanglement is demonstrated in the regime where the negative eigenvalues simultaneously exist for $l=1,2,3$ \cite{thrmode}. Also,  Fig.5~(d) shows that bipartite entanglement between the mode $c_2$ and the remaining two modes $c_{1}$ and $c_3$ is largest, since it is the only mode which is simultaneously entangled to the two other subsystems $c_1$ and $c_3$. Finally, the entanglement between the mode $c_3$ and the subsystem including $c_1$ and $c_2$ is smallest, since the bipartite entanglement satisfies $E_{23}<E_{12}$ and $E_{13}=0$. Therefore see that the negativities will satisfy $\lambda_{3-12}>\lambda_{1-23}>\lambda_{2-13}$.
\begin{figure}
\centerline{\scalebox{0.25}{\includegraphics{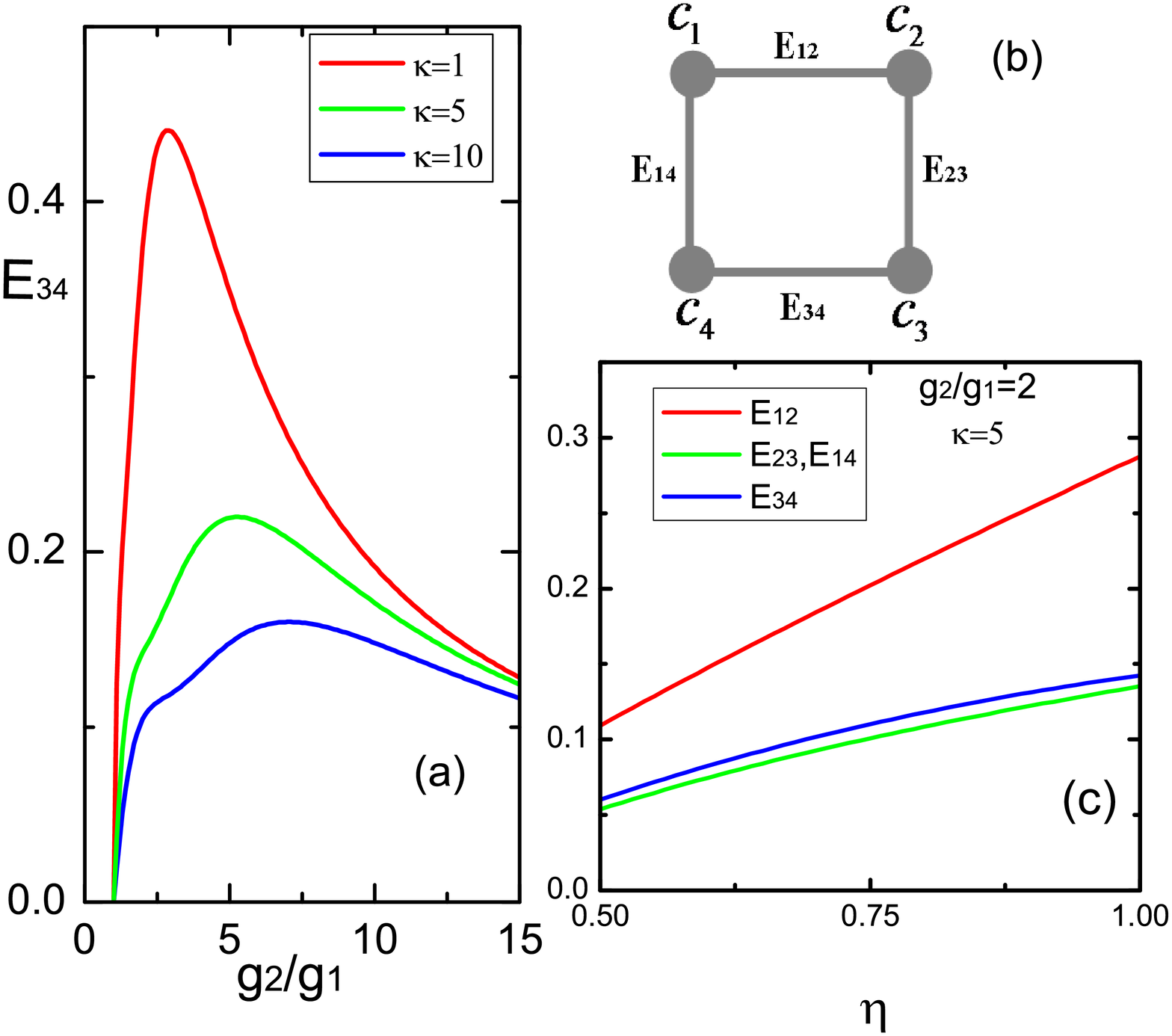}}} \caption{(a) Steady-state bipartite entanglement between the mechanical modes $c_3$ and $c_4$ for a four-mode mechanical system. (b) Quadripartite square graph-state
entanglement among the four remote mechanical oscillators. (c) Dependence of the reduced bipartite entanglement on the intercavity coupling efficiency $\eta$. Other parameters used are the same as in Fig.4.}
\end{figure}

When extending the above three-mode mechanical system to the four-mode case, i.e. $N=4$, it is not difficult to see from Eq.(\ref{nonadiab:lan4}) that the reduced bipartite entanglement $E_{12}$ and $E_{23}$ are not affected due to the unidirectional cavity coupling. Therefore,  the entanglement $E_{12}$ and $E_{23}$ are the the same as in the $N=3$ case plotted in Fig.5. Furthermore, it can be inferred from Eq.(\ref{adiab:lan4}) that the bipartite entanglements will satisfy $E_{14}=E_{23}$ in the bad-cavity limit.  In Fig.6~(a), we plot the bipartite entanglement $E_{34}$ between the mechanical modes $c_3$ and $c_4$, and it is obvious that it exhibits similar behavior as the entanglement $E_{12}$ between modes $c_1$ and $c_2$ ( see Fig.5). We therefore see that quadripartite square graph-state entanglement among four remote mechanical oscillators can be achieved via cascaded cavity couplings. This kind of multipartite entanglement is useful in the field of long-distance quantum communication.  The effects of imperfect cavity couplings are illustrated in Fig.6~(c). We see that for the coupling efficiency as low as $\eta=0.5$, the genuine quadripartite entanglement of four distant mechanical oscillators can still be achieved.

\section{Conclusion}
In conclusion, we propose a scheme to generate steady-state entanglement of remote mechanical oscillators in unidirectionally coupled cavities in the cascaded way. We note here that while the present model assumes the membranes as mechanical oscillators, the role of mechanical elements can also be played by momentum modes of clouds of ultracold atoms. By choosing the detuning of the pump lasers, in each cavity the mechanical oscillator is coupled to the two cavity modes via parametric and beam-splitter-like interactions. The output quantum fluctuating field of the first cavity is subsequently driving the second cavity with reversed detunings. For the case of two mechanical oscillators in cascaded cavities, the cavity dissipation can pull the two mechanical oscillators into a stationary two-mode squeezed vacuum state for negligible mechanical damping. The two-mode mechanical entanglement depends on the relative strength of the pump lasers and is robust to thermal fluctuations. For multiple mechanical oscillators in multiple cascaded cavities, it is found that the steady-state bipartite entanglement can be established between the odd and even oscillators, whereas odd and even oscillators do not become entangled. We show that using this scheme the genuine multipartite  entanglement can be achieved among remote mechanical oscillators by cavity dissipation. This kind of remote multipartite macroscopic entanglement is a useful resource in the construction of long-distance quantum communication networks.

\section*{Acknowledgment}
This work is supported by the National Natural Science
Foundation of China (Grant Nos. 11274134, 11074087, 61275123), the National Basic Research Program of
China (Grant No. 2012CB921602), the Natural Science Foundation of Hubei Province (Grant No. 2010CDA075), and the Natural Science Foundation of Wuhan City (Grant No. 201150530149). THT acknowledges the support from the CSC. BFB and HS acknowledge Pierre Meystre for his ongoing support.

\end{document}